%% file: ConferenceProceedings.tex
\RequirePackage{snapshot}
\documentclass{webofc}
\usepackage[varg]{txfonts}   
\usepackage{graphicx}
\usepackage{amsmath}
\usepackage{wrapfig}
\usepackage[caption=false]{subfig}
\usepackage{media9}

\usepackage{tikz,varwidth}
\usepackage{tikzscale}
\usetikzlibrary{shapes.geometric, patterns}
\usetikzlibrary{arrows.meta,shapes,automata,petri,positioning,calc,decorations.markings}
\usepackage{tkz-euclide}

\graphicspath{{./Figs/}}
\begin{document}
\title{Visualisations of Centre Vortices}
%
%

\author{
  \firstname{James} \lastname{Biddle}\inst{1}\fnsep\thanks{\email{james.biddle@adelaide.edu.au}}
  \and
  \firstname{Waseem} \lastname{Kamleh}\inst{1}\fnsep\thanks{\email{waseem.kamleh@adelaide.edu.au}}
  \and
  \firstname{Derek} \lastname{Leinweber}\inst{1}\fnsep\thanks{\email{derek.leinweber@adelaide.edu.au}}
}

\institute{Centre for the Subatomic Structure of Matter, Department of Physics, The
  University of Adelaide, SA 5005, Australia}

\abstract{
The centre vortex structure of the vacuum is visualised through the use of novel 3D visualisation techniques. These visualisations allow for a hands-on examination of the centre-vortex matter present in the QCD vacuum, and highlights some of the key features of the centre-vortex model. The connection between topological charge and singular points is also explored. This work highlights the useful role visualisations play in the exploration of the QCD vacuum.
}
\maketitle
\section{Introduction}
Our current understanding of the strong interaction is encapsulated in the gauge field
theory of Quantum Chromodynamics (QCD). Because the gauge bosons of QCD, the gluons, can
self-interact, the QCD vacuum is populated by highly non-trivial gluon and quark
condensates. However, it is not yet analytically determined what feature of the non-trivial QCD ground
state fields is fundamental to the distinctive properties of QCD, namely the
\begin{itemize}
\item Confinement of quarks, and
\item Dynamical chiral symmetry breaking leading to dynamical mass generation.
\end{itemize}
The most promising candidate supported by numerical studies is the centre vortex
picture~\cite{'tHooft:1977hy,'tHooft:1979uj}, which postulates that these two features are
caused by sheets of chromo-magnetic flux carrying charge associated with the centre of the
$SU(3)$ gauge group, given by the three values of $\sqrt[3]{1}$. The centre vortex picture
has already had much success in reproducing many distinctive QCD properties, such as the
linear static quark potential~\cite{Cais:2008za, Langfeld:2003ev,
  Trewartha:2015ida,Greensite:2003bk,DelDebbio:1998luz}, enhancement of the infrared gluon
propagator~\cite{Bowman:2010zr, Biddle:2018dtc, Langfeld:2001cz, Quandt:2010yq},
enhancement of the infrared quark mass function~\cite{Bowman:2008qd,Trewartha:2015nna} and
mass splitting in the low-lying hadron
spectrum~\cite{Trewartha:2017ive,Trewartha:2015nna,OMalley:2011aa}.\\

\begin{wrapfigure}{r}{0.4\linewidth}
  \centering
  \vspace{-10pt}
  \includegraphics[width=2.5cm]{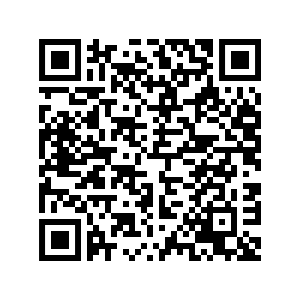}
  \vspace{-10pt}
  \caption{\label{fig:QRCode}QR code to download the augmented reality app for Android devices.}
\end{wrapfigure}

This work seeks to visualise these centre vortices on the lattice through the use of 3D
modelling techniques, allowing us to explore the vortex vacuum in a never-before-seen
way. These visualisations are presented as interactive 3D models embedded in the
document. To interact with these models, it is necessary to open the document in Adobe
Reader or Adobe Acrobat (requires version 9 or newer). Linux users should install Adobe
Acroread version 9.4.1, the last edition to have full 3D support. Note that 3D content
must also be enabled for the interactive content to be available, and for proper rendering
it is necessary to enable double-sided rendering in the preferences menu. To view the
models, click on the figures marked as \textbf{Interactive} in the caption. To rotate the
model, click and hold the left mouse button and move the mouse. Use the scroll wheel or
shift-click to zoom. Some pre-set views of the model are also provided to highlight areas
of interest. To reset the model back to its original orientation and zoom, press the
‘home’ icon in the toolbar or change the view to ‘Default view’. In addition, the 3D
models presented here can be viewed in augmented reality through Josh Charvetto's Android
application. See Fig~\ref{fig:QRCode} for a QR code link to download the app.

\section{Spatially-Oriented Centre Vortices}
When projected onto 3D space, vortices appear as closed lines carrying centre charge. They are
identified on the lattice by projecting the gluon field links onto their nearest centre
element in maximal centre gauge. Each $1\times 1$ Wilson loop, $P_{\mu\nu}(x)$, will then
take one of three possible values
\begin{equation}
  \label{eq:2}
  P_{\mu\nu}(x) = \exp\left(\frac{m 2\pi i}{3} \right)I,~ m\in \lbrace -1,\,0,\,+1 \rbrace.
\end{equation}
If $P_{\mu\nu}(x)$ takes one of the two complex phases, we say it is pierced by a
vortex. We refer to the vortex by it's centre charge parameter, $m=\pm 1$.\\

\begin{figure}
  \centering
  \scalebox{0.6}{\input{./Figs/SpaceVortices.tex}}
  \caption{\label{fig:SpacialVortices}An example of the plotting convention for vortices
    located within a 3D time slice. \textbf{Left:} A $+1$ vortex in the $+\hat{z}$
    direction. \textbf{Right:} A $-1$ vortex in the $-\hat{z}$ direction.}
\end{figure}
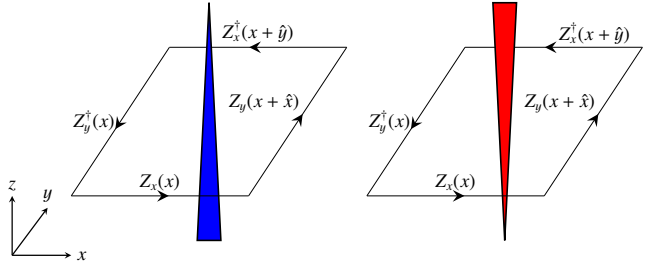

\begin{figure}[!h]
  \centering
  \includemedia[
  noplaybutton,
  3Dtoolbar,
  3Dmenu,
  label=Plaq_CFG95_T01.u3d,
  3Dviews=./Views/Views_Plaq_T01.vws,
  3Dcoo  = 10 10 20, 
  3Dc2c  = 0 1 0,    
  3Droo  = 50,       
  3Droll = 270,      
  3Dlights=CAD,
  width=0.8\linewidth,      
  ]{\includegraphics{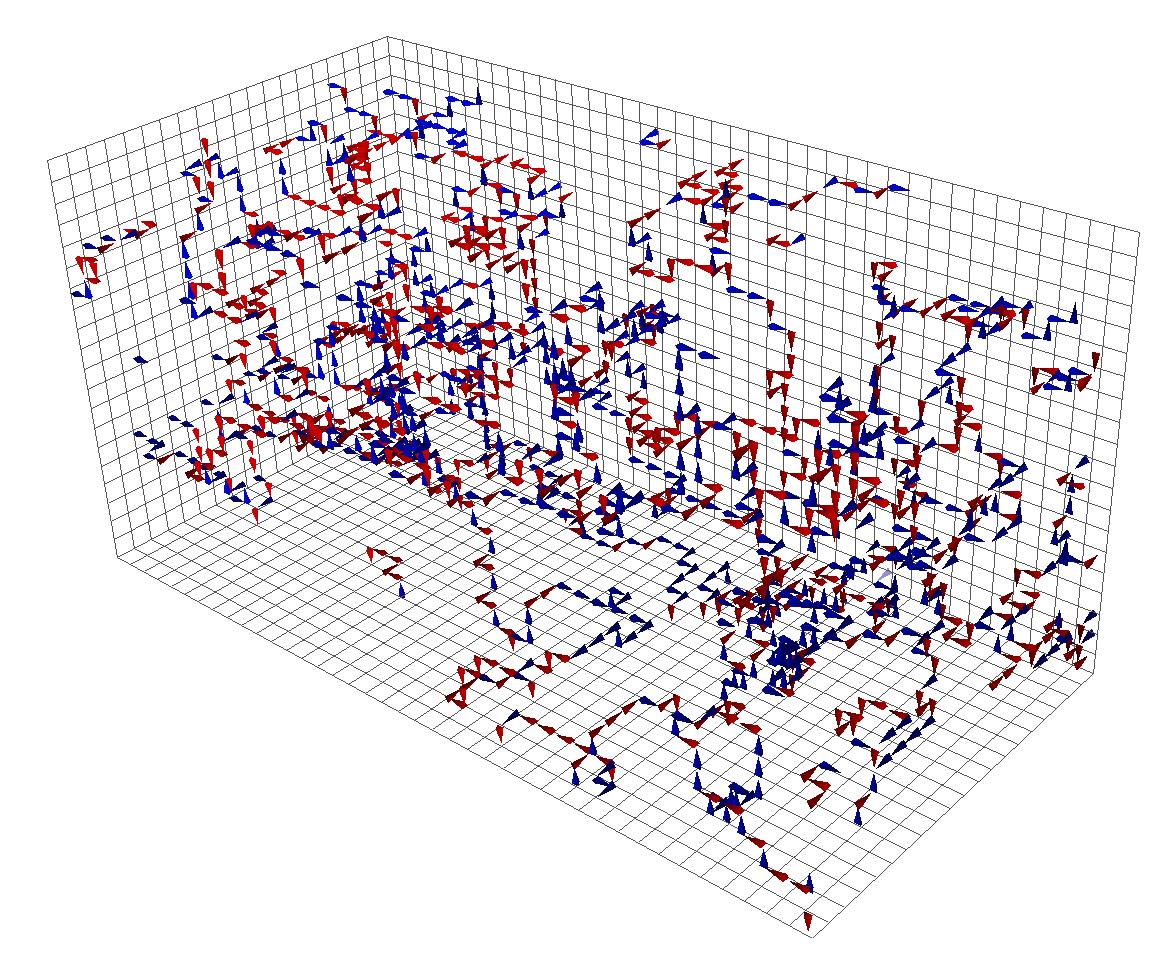}}{./U3D/Plaq_CFG95_T01.u3d}
  \caption{\label{fig:SpaceVortices} The first time-slice of spatially-oriented vortices. (\textbf{Interactive online})}
\end{figure}

\begin{figure}
  \centering
  \subfloat{
    \includegraphics[width=0.25\linewidth]{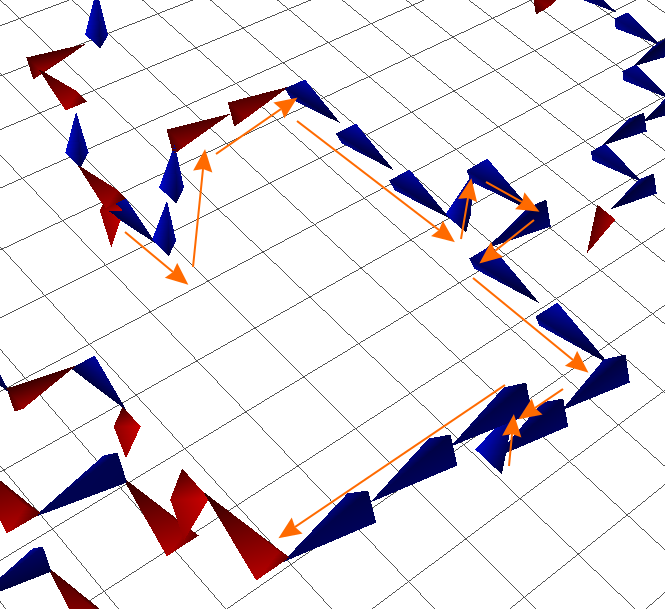}
  }\hfill \subfloat{
    \includegraphics[width=0.25\linewidth]{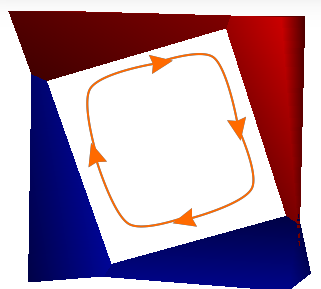}
  }\hfill \subfloat{
    \includegraphics[width=0.25\linewidth]{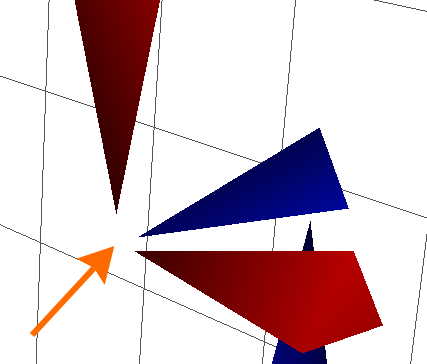}
  }
  \caption{\label{fig:VortexFeatures} \textbf{Left:} Vortices form continuous lines,
    highlighted with orange arrows in this diagram. \textbf{Middle:} Vortices must form
    closed loops to conserve the vortex flux. \textbf{Right:} $SU(3)$ vortices are capable
    of forming monopoles or branching points where three vortices emerge or converge at a
    single point.}
\end{figure}

For a charge $m=+1$ vortex, a blue jet is plotted positively oriented piercing the centre of the plaquette,
and for a charge $m=-1$ vortex, a red jet is plotted negatively oriented. An example of this convention is
shown in Fig.~\ref{fig:SpacialVortices}.
With these conventions, the first time slice of a gluon field configuration appears as
illustrated in Fig.~\ref{fig:SpaceVortices}. From this visualisation we note a couple of interesting properties.
Firstly, vortices must form closed lines to conserve centre flux. However, a vortex line is
permitted to branch into 2 lines due to the periodic property of the centre group. These
features are highlighted in Fig.~\ref{fig:VortexFeatures}. We also note that vortex loops tend to be large, which is indicative of the confining phase~\cite{Engelhardt:1999fd}.

\section{Space-Time Oriented Vortices}
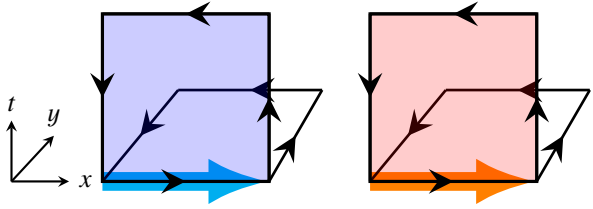
\begin{figure}
  \centering
  \subfloat{\tikzset{every picture/.style={scale=0.4}}
    \input{./Figs/p1TimeVortex.tex} }
  \subfloat{\tikzset{every picture/.style={scale=0.4}}
    \input{./Figs/m1TimeVortex.tex} }
  \caption{\textbf{Left:} A $+1$ vortex in the forward $x-t$ plane (shaded blue) will be
    indicated by a cyan arrow in the $+\hat{x}$ direction. \textbf{Right:} A $-1$ vortex
    in the forward $x-t$ plane (shaded red) will be indicated by an orange arrow in the
    $+\hat{x}$ direction.}
  \label{fig:TimeVortices}
\end{figure}

\begin{figure}[!h]
  \centering
  \includemedia[
  noplaybutton,
  3Dtoolbar,
  3Dmenu,
  label=PlaqLink_CFG95_T01.u3d,
  3Dviews=./Views/Views_PlaqLink_T01.vws,
  3Dcoo  = 10 10 20, 
  3Dc2c  = 0 1 0,    
  3Droo  = 50,       
  3Droll = 270,      
  3Dlights=CAD,
  width=0.8\linewidth,      
  ]{\includegraphics{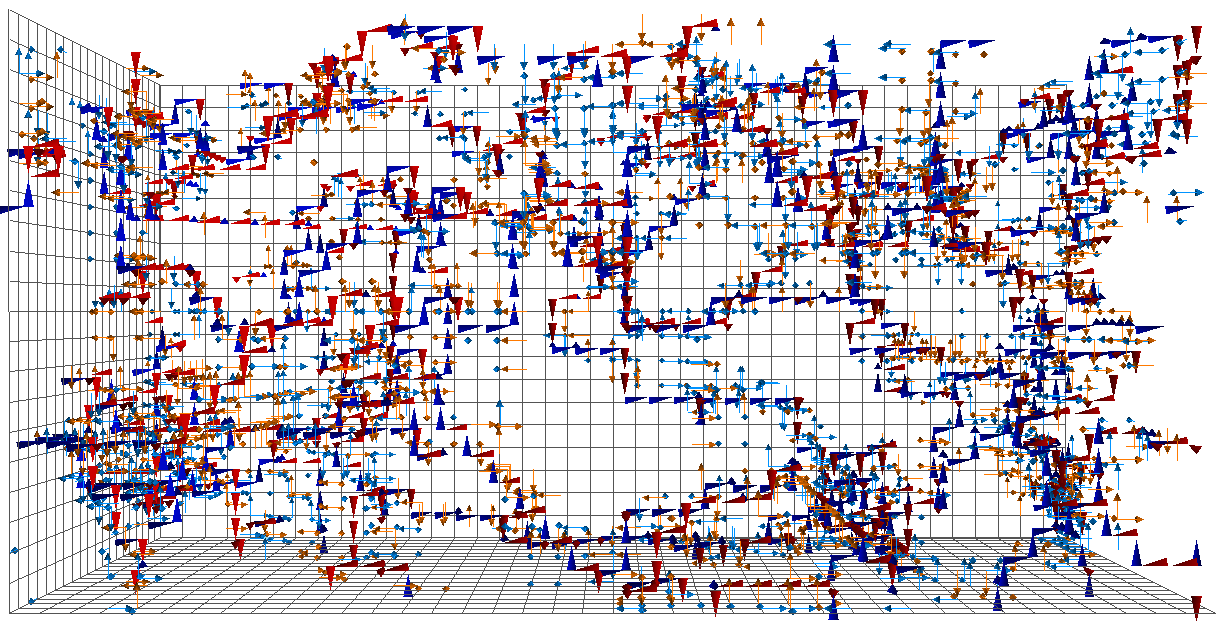}}{./U3D/PlaqLink_CFG95_T01.u3d}
  \caption{\label{fig:SpaceTimeVortices}The first time-slice of now containing both
    spatially-oriented and space-time vortices. (\textbf{Interactive online})}
\end{figure}

In 4D space-time, centre vortices map out a 2D wold sheet. To visualise these vortices, we
project onto 3D space where the sheets map out lines that vary with time. As we have taken
3D slices, we have suppressed all vortex information in the time direction. In each 3D
slice we only have access to one link belonging to the plaquettes associated with vortices
in the forwards and backwards $x_i - t$ planes. As such we plot an `indicator' link, to
signify the presence of a vortex in the suppressed direction. This follows the convention,
\begin{itemize}
\item \makebox[4cm]{$+1$ vortex, forward in time\hfill} $\implies$ cyan arrow,
  positively oriented
\item \makebox[4cm]{$+1$ vortex, backward in time\hfill} $\implies$ cyan arrow,
  negatively oriented
\item \makebox[4cm]{$-1$ vortex, forward in time\hfill}
  $\implies$ orange arrow, positively oriented
\item \makebox[4cm]{$-1$ vortex, backward in time\hfill}
  $\implies$ orange arrow, negatively oriented.
\end{itemize}
An example of these conventions is shown in Fig.~\ref{fig:TimeVortices}.
Adding these space-time indicator links to our previous visualisation, the first time
slice now appears as Fig.~\ref{fig:SpaceTimeVortices}.\\

\begin{figure}
  \centering
  \subfloat[\label{fig:VortexLadder1}$t=1$]{
    \includegraphics[height=0.5\linewidth]{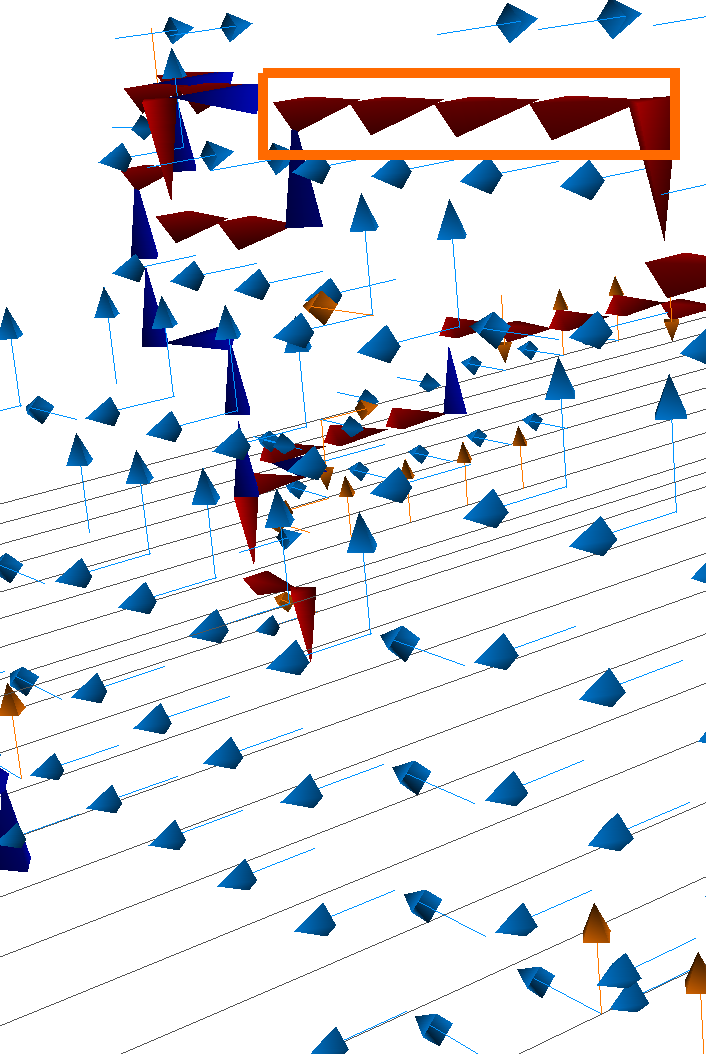}
  }
  \hspace{\columnsep}
  \subfloat[\label{fig:VortexLadder2}$t=2$]{
    \includegraphics[height=0.5\linewidth]{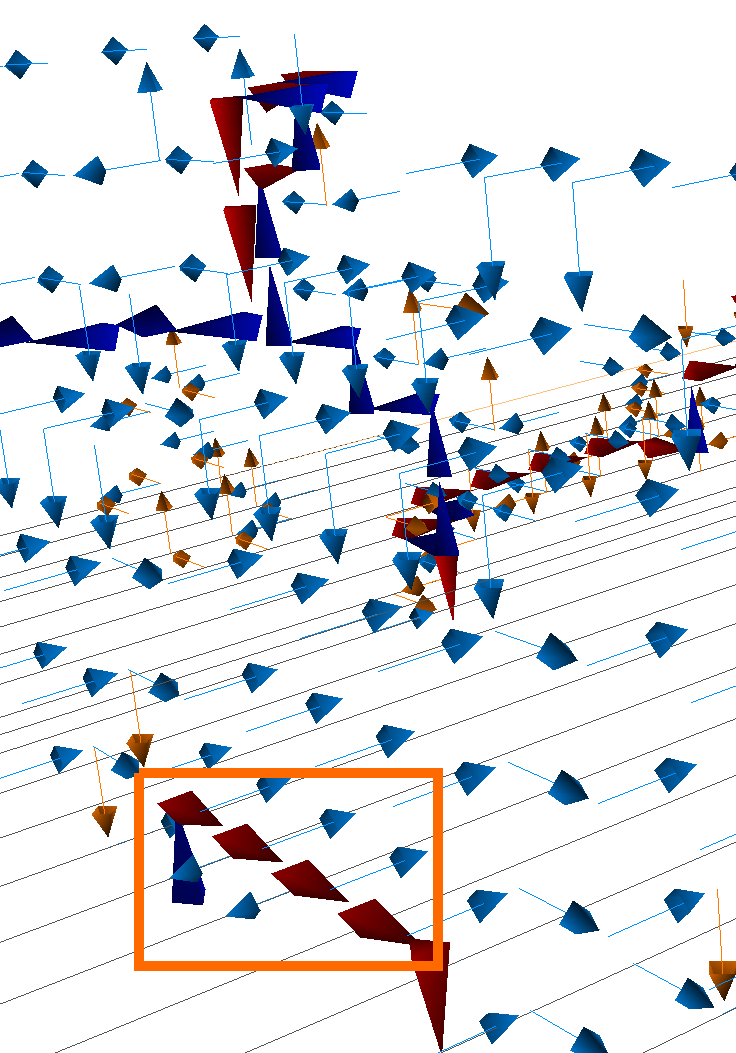}
  }
  \caption{\label{fig:VortexLadder}An example of a sheet of space-time oriented vortices
    predicting the motion of spatially-oriented vortices over multiple lattice sites
    from $t=1$ to $t=2$.}

\end{figure}

We can see how these space-time oriented indicator links predict the motion of
spatially-oriented vortices by looking at Fig.~\ref{fig:VortexLadder}. Here we see a line
of $m=-1$ vortices shifting along the sheet of cyan space-time oriented indicator links as
we step through time.

\section{Singular Points}
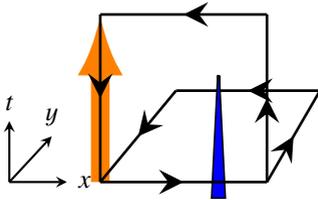
\begin{figure}
  \centering
  \begin{minipage}[c]{0.3\textwidth}
    \tikzset{every picture/.style={scale=0.4}}{\input{./Figs/SingularPoint.tex}}
  \end{minipage}\hfill
  \begin{minipage}[c]{0.67\textwidth}
    \caption{\label{fig:SingularPoint} The signature of a singular point, in which the
      tangent vectors of the vortex surface span all four dimensions. In this case, the
      blue jet is associated with field strength in the $x-y$ plane, and the orange
      space-time vortex indicator link is associated with a vortex generating field
      strength in the $z-t$ plane. Hence, the vortex surface spans all four dimensions at
      the front lower left corner of the illustration.}
  \end{minipage}
\end{figure}

\begin{figure}
  \centering
    \includemedia[
  noplaybutton,
  3Dtoolbar,
  3Dmenu,
  label=PlaqLink_CFG95_T01.u3d,
  3Dviews=./Views/Views_PlaqLink_T01.vws,
  3Dcoo  = 10 10 20, 
  3Dc2c  = 0 1 0,    
  3Droo  = 50,       
  3Droll = 270,      
  3Dlights=CAD,
  width=0.8\linewidth,      
 ]{\includegraphics{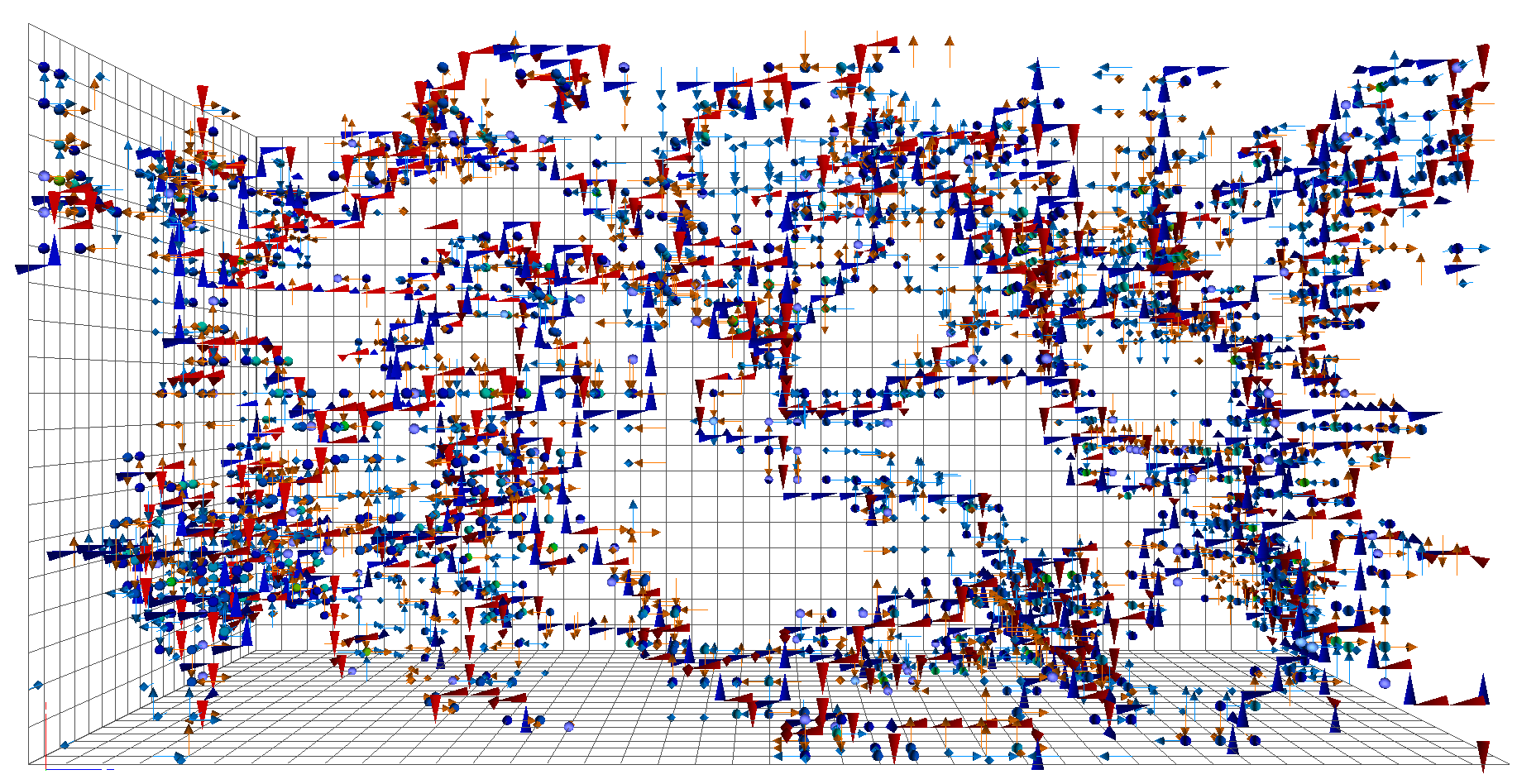}}{./U3D/VortexVis_CFG95_NoCool_Plaq_Link_SingularPtCount.u3d}
 \caption{\label{fig:SPts}The singular points on the $t=1$ time slice presented in
   Fig.~\ref{fig:SpaceTimeVortices} are identified by spheres. (\textbf{Interactive online})}
\end{figure}

Using our vortex illustrations, we can identify locations where vortex surfaces span all
four space-time dimensions. These locations are known as singular points, and can be
identified as shown below by an indicator link running parallel to a spatially-oriented
jet, as shown in Fig.~\ref{fig:SingularPoint}.
These points are significant because they necessarily generate regions of topological
charge density. A visualisation of these singular points is presented in Fig.~\ref{fig:SPts}.
\clearpage
\section{Conclusion}
Visualisations of centre vortices provide valuable insight into the nature of the QCD
vacuum. Through these visualisations we can identify structures of interest such as
branching points and singular points, and study their relationship with topological charge
density. For a more detailed description of these visualisations, and an analysis of the
correlation between vortices and topological charge, see Ref.~\cite{Biddle:2019gke}. Work
such as this allows one to explore QCD and centre vortices in a novel manner, and provides
exciting new perspectives on the centre vortex vacuum.
\section{Acknowledgments}
The authors wish to thank Amalie Trewartha for her contributions to the gauge ensembles
underlying this investigation. We also thank Ian Curington of Visual Technology Services
Ltd. for his support of the PDF3D software used in creating the interactive 3D
visualisations presented herein. This research is supported with supercomputing resources
provided by the Phoenix HPC service at the University of Adelaide and the National
Computational Infrastructure (NCI) supported by the Australian Government. This research
is supported by the Australian Research Council through Grants No. DP190102215,
DP150103164, DP190100297 and LE190100021.

\bibliography{ConferenceProceedings}
      %
      %
      %
      %

\end{document}

%% file: Figs/SpaceVortices.tex
\begin{tikzpicture}[scale=1.3]
\tkzDefPoint(0,0){BL1}
\tkzDefPoint(1.66,2.5){TL1}
\tkzDefShiftPoint[TL1](3,0){TR1}
\tkzDefShiftPoint[BL1](3,0){BR1}

\tkzDefPoint(2.33,1.25){mid1}
\tkzDefShiftPoint[mid1](-0.2,-2){triBL1}
\tkzDefShiftPoint[mid1](0.2,-2){triBR1}
\tkzDefShiftPoint[mid1](0,2){triT1}

\tkzDefPoint(-1,-1){ax}
\tkzDefShiftPoint[ax](1,0){axR}
\tkzDefShiftPoint[ax](0,1){axT}
\tkzDefShiftPoint[ax](0.6,0.8){axB}

\begin{scope}[very thick,decoration={
    markings,
    mark=at position 0.55 with {\arrow[scale=2]{stealth}}}
    ] 

	\tkzDrawSegments[postaction={decorate}](TR1,TL1)
	\filldraw[fill=blue, line width=1pt](triT1) -- (triBL1) -- (triBR1) -- cycle;
	\tkzDrawSegments[postaction={decorate}](BR1,TR1 BL1,BR1 TL1,BL1); 
    
\end{scope}

\tkzLabelSegment(BL1,BR1){\large $Z_x(x)$}
\tkzLabelSegment[left](TL1,BL1){\large $Z^\dagger_y(x)$}
\tkzLabelSegment[label={[label distance=-0.3cm]160:\large $Z_y(x+\hat{x})$}](BR1,TR1){}
\tkzLabelSegment[above](TR1,TL1){\large $Z_x^\dagger(x+\hat{y})$}

\tkzDrawSegments[thick,->, >=stealth](ax,axR)
\tkzDrawSegments[thick,->, >=stealth](ax,axT)
\tkzDrawSegments[thick,->, >=stealth](ax,axB)
\tkzLabelPoint[right](axR){\large $x$}
\tkzLabelPoint[above](axT){\large $z$}
\tkzLabelPoint[above](axB){\large $y$}

\tkzDefPoint(5,0){BL2}
\tkzDefPoint(6.66,2.5){TL2}
\tkzDefShiftPoint[TL2](3,0){TR2}
\tkzDefShiftPoint[BL2](3,0){BR2}

\tkzDefPoint(7.33,1.25){mid2}
\tkzDefShiftPoint[mid2](-0.2,2){triBL2}
\tkzDefShiftPoint[mid2](0.2,2){triBR2}
\tkzDefShiftPoint[mid2](0,-2){triT2}

\begin{scope}[very thick,decoration={
    markings,
    mark=at position 0.55 with {\arrow[scale=2]{stealth}}}
    ] 

	\tkzDrawSegments[postaction={decorate}](TR2,TL2)
	\filldraw[fill=red, line width=1pt](triT2) -- (triBL2) -- (triBR2) -- cycle;
	\tkzDrawSegments[postaction={decorate}](BR2,TR2 BL2,BR2 TL2,BL2);    
    
\end{scope}
\tkzLabelSegment(BL2,BR2){\large $Z_x(x)$}
\tkzLabelSegment[left](TL2,BL2){\large $Z^\dagger_y(x)$}
\tkzLabelSegment[label={[label distance=-0.3cm]160:\large $Z_y(x+\hat{x})$}](BR2,TR2){}
\tkzLabelSegment[label={[label distance=-0.3cm]87:\large $Z_x^\dagger(x+\hat{y})$}](TR2,TL2){}
\end{tikzpicture}

%% file: Figs/p1TimeVortex.tex
\begin{tikzpicture}
  \tkzDefPoint(-7,-1.5){ax}
  \tkzDefShiftPoint[ax](1.9,0){axR}
  \tkzDefShiftPoint[ax](0,2){axT}
  \tkzDefShiftPoint[ax](1.4,1.5){axB}

  \begin{scope}[very thick,decoration={
      markings,
      mark=at position 0.5 with {\arrow[scale=2]{stealth}}}
    ] 

    \draw[line width=7,color=cyan,-{Latex[length=8mm]}](-4.0,-1.5)--(1.5,-1.5);

    \draw[line width=1.0,postaction={decorate}](1.5,-1.5)--node[right]{} (3.25,1.5)node(g){};
    \draw[line width=1.0,postaction={decorate}](3.25,1.5)--node[above]{} (-1.5,1.5);
    \draw[line width=1.0,postaction={decorate}](-1.5,1.5)--node[left]{} (-4,-1.5);

    \filldraw[fill=blue,fill opacity=0.2] (-4,-1.5) -- (1.5,-1.5) -- (1.5,4) -- (-4,4)--cycle;

    
    \draw[line width=1.0,postaction={decorate}](-4.0,-1.5)--node[above]{}(1.5,-1.5)node(f){};
    
    \draw[line width=1.0,postaction={decorate}]( 1.5,-1.5)--node[right]{} (1.5,4.0);    
    \draw[line width=1.0,postaction={decorate}]( 1.5, 4.0)--node[above]{}(-4.0,4.0);    
    \draw[line width=1.0,postaction={decorate}](-4.0, 4.0)--node[left]{}(-4.0,-1.5);

  \end{scope}
  \tkzDrawSegments[thick,->, >=stealth](ax,axR)
  \tkzDrawSegments[thick,->, >=stealth](ax,axT)
  \tkzDrawSegments[thick,->, >=stealth](ax,axB)
  \tkzLabelPoint[right](axR){$x$}
  \tkzLabelPoint[above](axT){$t$}
  \tkzLabelPoint[above](axB){$y$}  
\end{tikzpicture}

%% file: Figs/m1TimeVortex.tex
\begin{tikzpicture}
\begin{scope}[very thick,decoration={
    markings,
    mark=at position 0.5 with {\arrow[scale=2]{stealth}}}
    ] 

  \draw[line width=7,color=orange,-{Latex[length=8mm]}](-4.0,-1.5)--(1.5,-1.5);

  \draw[line width=1.0,postaction={decorate}](1.5,-1.5)--node[right]{} (3.25,1.5)node(g){};
  \draw[line width=1.0,postaction={decorate}](3.25,1.5)--node[above]{} (-1.5,1.5);
  \draw[line width=1.0,postaction={decorate}](-1.5,1.5)--node[left]{} (-4,-1.5);

    \filldraw[fill=red,fill opacity=0.2] (-4,-1.5) -- (1.5,-1.5) -- (1.5,4) -- (-4,4)--cycle;
  
  
  \draw[line width=1.0,postaction={decorate}](-4.0,-1.5)--node[above]{}(1.5,-1.5)node(f){};
  
  \draw[line width=1.0,postaction={decorate}]( 1.5,-1.5)--node[right]{} (1.5,4.0);    
  \draw[line width=1.0,postaction={decorate}]( 1.5, 4.0)--node[above]{}(-4.0,4.0);    
  \draw[line width=1.0,postaction={decorate}](-4.0, 4.0)--node[left]{}(-4.0,-1.5);

  \end{scope}
\end{tikzpicture}

%% file: Figs/SingularPoint.tex
\begin{tikzpicture}
  \tkzDefPoint(-7,-1.5){ax}
  \tkzDefShiftPoint[ax](1.9,0){axR}
  \tkzDefShiftPoint[ax](0,2){axT}
  \tkzDefShiftPoint[ax](1.4,1.5){axB}

  \begin{scope}[very thick,decoration={
      markings,
      mark=at position 0.5 with {\arrow[scale=2]{stealth}}}
    ] 

    \draw[line width=7,color=orange,-{Latex[length=8mm]}](-4.0,-1.5)--(-4.0,4.0);

    \draw[line width=1.0,postaction={decorate}](1.5,-1.5)--node[right]{} (3.25,1.5)node(g){};
    \draw[line width=1.0,postaction={decorate}](3.25,1.5)--node[above]{} (-1.5,1.5);
    \draw[line width=1.0,postaction={decorate}](-1.5,1.5)--node[left]{} (-4,-1.5);

    \draw (-0.3,-2) node(a){}
    -- (0.1,-2) node(b){}   
    -- (-0.1,2) node(c){}   
    -- cycle;               
    \fill[blue] (a.center) -- (b.center) -- (c.center);
    
    \draw[line width=1.0,postaction={decorate}](-4,-1.5)--node[above]{}(1.5,-1.5)node(f){};
    
    \draw[line width=1.0,postaction={decorate}]( 1.5,-1.5)--node[right]{} (1.5,4.0);    
    \draw[line width=1.0,postaction={decorate}]( 1.5, 4.0)--node[above]{}(-4.0,4.0);    
    \draw[line width=1.0,postaction={decorate}](-4.0, 4.0)--node[left]{}(-4.0,-1.5);

  \tkzDrawSegments[thick,->, >=stealth](ax,axR)
  \tkzDrawSegments[thick,->, >=stealth](ax,axT)
  \tkzDrawSegments[thick,->, >=stealth](ax,axB)
  \tkzLabelPoint[right](axR){$x$}
  \tkzLabelPoint[above](axT){$t$}
  \tkzLabelPoint[above](axB){$y$}  
  \end{scope}
\end{tikzpicture}